%% file: main.tex
\documentclass[conference]{IEEEtran}
\IEEEoverridecommandlockouts
\usepackage{cite}
\usepackage{amsmath,amssymb,amsfonts}
\usepackage{graphicx}
\usepackage{textcomp}
\usepackage{comment}
\usepackage{orcidlink}
\usepackage[dvipsnames]{xcolor}
\usepackage{enumitem}
\usepackage{pifont}
\usepackage{makecell}
\usepackage{listings}
\usepackage[normalem]{ulem}
\usepackage{adjustbox}
\usepackage{multirow}
\usepackage{subfig}

\usepackage{algorithm}
\usepackage{algpseudocode}

\renewcommand{\checkmark}{\textcolor{ForestGreen}{\ding{51}}}
\newcommand{\crossmark}{\textcolor{Red}{\ding{55}}}

\newcommand{\name}{\mbox{\textsc{Pasta}}}

\newcommand{\rev}[1]{\begingroup\color{black}#1\endgroup}

\newcommand{\rephrase}[1]{\begingroup\color{black}#1\endgroup}

\newcommand{\hyeran}[1]{\textcolor{brown}{#1}}

\usepackage{xstring}
\usepackage{xurl}       
\def\UrlBreaks{\do\_\do\:} 
\makeatletter
\newcommand{\textttbreak}[1]{%
  \begingroup
    \StrSubstitute{#1}{\_}{_}[\ttb@tmp]%
    \expandafter\endgroup
    \expandafter\path\expandafter{\ttb@tmp}%
}
\makeatother
\let\origtexttt\texttt
\renewcommand{\texttt}[1]{\textttbreak{#1}}

\definecolor{darkgreen}{rgb}{0.0, 0.5, 0.0}
\def\BibTeX{{\rm B\kern-.05em{\sc i\kern-.025em b}\kern-.08em
    T\kern-.1667em\lower.7ex\hbox{E}\kern-.125emX}}
\begin{document}

\title{PASTA: A Modular Program Analysis Tool Framework for Accelerators}
\author{\IEEEauthorblockN{
    Mao Lin\IEEEauthorrefmark{1}\orcidlink{0000-0002-1460-0766},
    Hyeran Jeon\IEEEauthorrefmark{1}\orcidlink{0000-0002-1767-8198},
    Keren Zhou\IEEEauthorrefmark{2}\IEEEauthorrefmark{3}\orcidlink{0000-0002-7977-3182},
}\\
  \IEEEauthorblockN{
  \IEEEauthorrefmark{1}University of California, Merced, USA \
  \IEEEauthorrefmark{2}George Mason University, USA \
  \IEEEauthorrefmark{3}OpenAI, USA}
}
\maketitle
\begingroup
\renewcommand\thefootnote{}
\footnotetext{This paper has been accepted to the IEEE/ACM International Symposium on Code Generation and Optimization (CGO), 2026.}
\endgroup
\begin{abstract}
The increasing complexity and diversity of hardware accelerators in modern computing systems demand flexible, low-overhead program analysis tools. We present \name{}, a low-overhead and modular \underline{P}rogram \underline{A}nalysi\underline{S} \underline{T}ool Framework for \underline{A}ccelerators. \name{} abstracts over low-level profiling APIs and diverse deep learning frameworks, offering users a unified interface to capture and analyze runtime events at multiple levels. Its extensible design enables researchers and practitioners to rapidly prototype custom tools with minimal overhead. We demonstrate the utility of \name{} by developing several analysis tools, including tools for 
deep learning workload characterization 
and 
UVM optimization. 
Through extensive evaluations on mainstream deep learning workloads \rev{tested on 
NVIDIA and AMD GPUs under 
both single- and multi-GPU scenarios, we demonstrate \name{}’s broad applicability. On NVIDIA GPUs, we further} show that \name{} provides detailed performance insights with significantly lower overhead (up to 1.3$\times$10\textsuperscript{4} 
faster) than conventional analysis tools, thanks to its GPU-accelerated backend. \name{} strikes a practical balance between usability, extensibility, and efficiency, making it well-suited for modern accelerator-based computing environments.

\end{abstract}

\begin{IEEEkeywords}
Performance Analysis, Performance Tool, GPU Computing, Deep Learning, NVIDIA GPU, AMD GPU, Heterogeneous Accelerator
\end{IEEEkeywords}

%
%

\input{introduction}

\input{background}
\input{design}
\input{implementation}
\input{case-studies}
\input{discussion}
\input{conclusion}

\input{appendix}

\newpage

{
  \makeatletter
  \def\UrlBreaks{\do\/\do\.\do\-\do\_} 
  \makeatother
  \urlstyle{same}                      

  \let\texttt\origtexttt

  \bibliographystyle{IEEEtran}
  \bibliography{references}
}

\end{document}

%% file: introduction.tex
\section{Introduction
}
\begin{table*}[t]
\caption{Comparison of \name{} with tools from accelerator vendors and deep learning frameworks. 
}
\centering
\scriptsize
\resizebox{0.85\linewidth}{!}{
\begin{tabular}{|c|c|c|c|c|c|c|}
\hline
Tool/Fuctionalities & \makecell{NVIDIA\\Supported$^1$} & \makecell{AMD\\Supported$^2$} & \makecell{DL Framework\\Supported$^3$} & \makecell{Low-overhead\\(GPU-accelerated)} & \makecell{Extensibility} & \makecell{Open-\\Sourced}\\
\hline
\hline
\name{} (Ours) & \checkmark & \checkmark & \checkmark & \checkmark & \checkmark  & \checkmark \\
\hline
NSight Systems~\cite{nvidia-systems} & \checkmark & \crossmark & \crossmark & \crossmark & \crossmark & \crossmark \\
\hline
ROCProfiler~\cite{rocprofiler} & \crossmark & \checkmark & \crossmark & \crossmark & \crossmark & \checkmark \\
\hline
PyTorch Profiler~\cite{pytorch-profiler} & \crossmark & \crossmark & \checkmark & \crossmark & \crossmark & \checkmark \\
\hline
TensorFlow Profiler~\cite{tensorflow-profiler} & \crossmark & \crossmark & \checkmark & \crossmark & \crossmark & \checkmark \\
\hline
Omniperf~\cite{omniperf} & \crossmark & \checkmark & \crossmark & \crossmark & \crossmark & \checkmark \\
\hline
\multicolumn{7}{l}{\tiny {\vbox to 3ex{\vfil}} $^{1, 2}$: Tools can analyze standard NVIDIA CUDA and AMD ROCm programs with low-level vendor library information.}\\
\multicolumn{7}{l}{\tiny {\vbox to 2ex{\vfil}} $^{3}$: Tools can capture and analyze deep learning framework-specific events, such as tensor allocation and destruction}\\
\end{tabular}
}\vspace{-12pt}
\label{tab:tools}
\end{table*}

\rephrase{With Moore’s Law nearing its physical limits, the escalating computational needs of emerging big data workloads have ushered in the era of domain-specific computing. 
Various accelerators, such as GPUs and TPUs, have emerged as essential compute engines 
with their massive parallelism and specialized compute capabilities. 
}
To fully exploit the compute capabilities of these accelerators, understanding workload behavior and identifying performance bottlenecks are crucial.
However, the massive parallelism within individual accelerators, combined with their asynchronous interactions with CPUs, complicates the task of performance analysis and hinders the deduction of actionable optimization insights.
To address this, accelerator vendors offer performance analysis tools such as NVIDIA Nsight Systems~\cite{nvidia-systems} and AMD ROCm Profiler~\cite{rocprofiler}, which help developers analyze execution behavior and performance metrics.
Despite their usefulness, these vendor-provided tools have notable limitations including \emph{limited flexibility and extensibility} and \emph{inadequate support for emerging workloads}.
These tools typically focus on predefined general-purpose metrics and often fail to meet the needs of user-specific performance analysis. For example, NVIDIA Nsight Systems provides timeline-based views of CPU and GPU activity, but it lacks the ability to capture fine-grained memory reuse patterns or associate memory activity with high-level model structures in deep learning (DL) workloads.
Furthermore, vendor tools primarily capture events from low-level profiling libraries~\cite{nvidia-cupti, rocprofiler, intel-oneapi} and attribute them to general program context, offering limited visibility into higher-level behaviors specific to modern DL frameworks like PyTorch~\cite{paszke2019pytorch} and TensorFlow~\cite{abadi2016tensorflow}.
For example, these frameworks introduce their own memory allocators and execution models (e.g., kernel grouping by layers), which result in unique runtime patterns and performance characteristics.
While DL frameworks integrate their own profilers, such as the PyTorch Profiler~\cite{pytorch-profiler} and TensorFlow Profiler~\cite{tensorflow-profiler}, these profilers only expose high-level model behaviors and lack the capability to trace low-level accelerator performance details and are not easily extensible.

To support custom performance analysis, accelerator vendors also expose low-level programming interfaces and libraries~\cite{compute-santizer-api, villa2019nvbit, rocprofiler-sdk, tpu_execution_profiler}, enabling developers to build performance analysis tools tailored to their specific needs. However, this approach often requires comprehensive knowledge of accelerator architecture, low-level programming models, as well as considerable development effort. 
In addition, while various community-developed tools exist for accelerator performance analysis, they only target specific inefficiency problems or specialized use cases~\cite{lin2023drgpum, zhou2020gvprof, zhao2024deep, nayak2024over, you2024gvarp}.
Their limited extensibility poses challenges for generalization to a broader range of program analysis tasks.

Given the increasing demand for performance analysis tools tailored to emerging workloads and the limitations of existing solutions, we present \textit{\name{}}, a low-overhead modular program analysis tool 
framework for accelerators. To the best of our knowledge, \name{} is the first framework designed to support cross-vendor accelerators and diverse DL workloads with extensibility. \name{} offers several key advantages over existing solutions.
\textbf{1) Modularity and extensibility.} \name{} can be easily extended to meet user-specific performance analysis needs by allowing developers to create custom analyses with simply overriding functions in the \name{} tool collection template.
\textbf{2) Cross-vendor support.} \name{} abstracts away the differences among vendor-specific profiling interfaces through unified \name{} event handlers, which support 
monitoring on various accelerator architectures. 
\textbf{3) DL framework integration.} \name{} supports DL framework-specific events 
capturing functionalities (e.g., operator execution and tensor allocation) to provide a more holistic view of workload behavior.
\textbf{4) Low-overhead design.} 
\name{} is designed with analysis efficiency in mind, aiming to reduce runtime impact and accelerate the processing of performance data. It leverages lightweight hooks provided by vendor profiling interfaces and DL framework callbacks, minimizing instrumentation overhead.
Additionally, \name{} includes an event processor that preprocesses raw runtime data and performs preliminary analysis on GPUs, accelerating the processing of large volumes of data generated by massively parallel accelerator executions. 
Table~\ref{tab:tools} compares the key advantages of \name{} with the tools provided by accelerator vendors and DL frameworks.

To demonstrate the practical utility of \name{}, we develop several analysis tools as case studies, including a DL workload characterization tool and a \rev{Unified Virtual Memory} \rev{(}UVM\rev{)} optimization tool.
These tools are implemented with minimal effort thanks to \name{}’s extensible framework and can be applied to diverse DL models.
The case studies reveal actionable insights, such as identifying 
kernel bottlenecks, quantifying underutilized memory regions, and optimizing UVM prefetching strategies, with 
significantly lower overhead compared to existing profiling tools, particularly due to \name{}’s GPU-accelerated analysis design.

Our contributions are as follows:
\begin{enumerate}[label=\textbullet, leftmargin=*, labelindent=2pt]
\item To the best of our knowledge, 
\name{} is the first 
program analysis 
framework that supports emerging DL workloads and accelerators from multiple vendors\rev{, including both NVIDIA and AMD GPUs}.

\item \name{}'s 
modular and extensible design allows it 
to be easily tailored 
to diverse 
performance analysis needs, which significantly speeds up the optimization and development processes. 

\item \rev{We demonstrate the effectiveness of \name{} through case studies on single- and multi-GPU scenarios}, which demonstrate substantially reduced analysis time and unique insights with \name{} compared with the existing vendor-specific profiling tools.  

\item \name{} is fully open-sourced under the MIT license\footnote{Source code available at: \textcolor{BlueViolet}{\url{https://github.com/AccelProf/AccelProf}}} and includes a detailed user guide for conducting performance analysis with some tools built using \name{}, as well as a 
developer guide for extending \name{} for their own performance analysis needs\footnote{User and developer documentation: \textcolor{BlueViolet}{\url{https://accelprofdocs.readthedocs.io}}}.
\end{enumerate}

%% file: background.tex
\section{Background and Related Work
}
This section uses GPUs as an example accelerator, while \name{} can be used for any accelerators 
\rephrase{that have APIs with which a host CPU can monitor various execution status}. 

\subsection{GPU Performance Analysis}
\rephrase{
Due to massive parallelism and asynchronous interactions with a host CPU, GPU-accelerated applications pose challenges for performance analysis.}
\rephrase{To support analysis,} GPU vendors provide various tools such as NVIDIA's Nsight Systems~\cite{nvidia-systems} and Nsight Compute~\cite{nvidia-compute}, AMD's ROC Profiler~\cite{rocprofiler} and Omniperf~\cite{omniperf}, and Intel’s VTune Profiler~\cite{intel-vtune}. 
\rephrase{These tools profile low-level activities of architectural components, but those profiling results often lack application semantics 
and insufficient to provide meaningful insights for optimization, 
making them difficult to use directly for performance tuning.}

To address these limitations, 
several 
analysis tools 
have been developed using vendor-provided interfaces.
DrGPUM\cite{lin2023drgpum} and Diogenes\cite{welton2019diogenes} pinpoint memory-related inefficiencies, such as inefficient CPU-GPU memory transfers.
ValueExpert and GVProf\cite{zhou2022valueexpert, zhou2020gvprof} identify value-related inefficiencies in GPU-accelerated applications. 
Nayak et al. proposed a tool that identifies redundant and improper synchronization operations in GPU programs\cite{nayak2024over}. GVARP\cite{you2024gvarp} detects performance variance in large-scale heterogeneous systems and provides insights to locate the root causes. CUDAAdvisor~\cite{shen2018cudaadvisor} performs fine-grained GPU kernel analysis, including memory reuse distance and divergence analysis, offering actionable insights for optimization.
While these tools enable more comprehensive performance analysis, as each tool is designed for certain inefficiencies of the target GPUs, users should identify the right tools for each target analysis. 
Some tools, such as HPCToolkit~\cite{adhianto2010hpctoolkit} supports a more general performance analysis on both NVIDIA and AMD GPUs. However, they primarily focus on HPC workloads and do not 
support 
emerging 
workloads such as DL models. 
\rephrase{Given the diversity of accelerators and workloads, a more extensible solution is needed to support broader analyses.}

\subsection{DL Workload Performance Analysis}
With the increasing importance of DL models for almost all computing domains, performance optimization of DL workloads is one of the most critical research topics of today. 
However, most of the DL frameworks are designed to be DL practioner-friendly 
while hiding backend interactions with accelerators
~\cite{ml-frameworks1, ml-frameworks2}. 
\rephrase{While this abstraction helps DL model designers to focus on the model architecture without concerning systems and hardware-side activities, 
it complicates execution analysis and optimization using traditional performance tools.}

To tackle this issue, DL frameworks have introduced their own performance analysis tools, such as 
PyTorch Profiler~\cite{pytorch-profiler}, 
TensorFlow Profiler~\cite{tensorflow-profiler} and JAX Profiler~\cite{jax-profiler}. 
While these framework-native performance analysis tools are useful, they have several limitations: (1) they 
lack support for exposing low-level details of accelerators, (2) they often require significant programming effort to 
configure, and (3) they support limited 
profiling metrics, which lack 
flexibility and extensibility for customized analysis.

To overcome these limitations, 
third-party DL performance analysis tools have been introduced. NVIDIA provides DLProf~\cite{dlprof}, which aggregates kernel performance data from tools like Nsight Systems and nvprof and offers layer-wise kernel performance summaries. DeepContext~\cite{zhao2024deep} links call stacks from high-level Python code to underlying accelerator C/C++ libraries, 
enabling the identification of inefficiencies in DL codebases. RL-Scope~\cite{gleeson2021rl} collects cross-stack profiling information (e.g., CUDA API time and GPU kernel time) and provides a detailed breakdown of CPU/GPU execution time. Hotline Profiler~\cite{snider2023hotline} detects runtime bottlenecks in DL workloads and presents them using multi-scale timeline visualizations.
\rephrase{Despite these advances, these tools either support only specific target inefficiencies or remain closed-source, thus are difficult to extend for customized analysis across diverse DL workloads.}

%% file: design.tex
\section{Design and Methodology
}

\begin{figure}[t]
    \centering
    \includegraphics[width=0.85\linewidth]{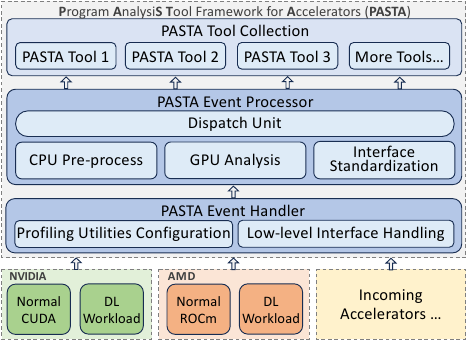}
    \caption{Design of \name{}.}
    \Description{}
    \label{fig:design}\vspace{-12pt}
\end{figure}

\subsection{Overall Design}
Figure~\ref{fig:design} shows the architecture of \name{}, which consists of three modular components: \emph{\name{} Event Handler}, \emph{\name{} Event Processor}, and \emph{\name{} Tool Collection}.
The event handler interfaces directly with low-level, vendor-specific profiling APIs and high-level DL framework callbacks to configure. 
This layer abstracts away the complexity of diverse accelerator platforms and enables consistent event collection across hardware vendors.
Built atop the event handler, the event processor acts as the dispatch and preprocessing layer.
It standardizes heterogeneous runtime information through a unified interface and performs preprocessing on either the CPU or GPU. 
This component transforms raw profiling data into structured insights suitable for higher-level analysis.
\rephrase{The tool collection hosts user-defined analysis tools that retrieve runtime data via the standardized interface and perform customized analyses such as kernel profiling or memory characterization.}
All three components are designed in independent modules so that each can be separately upgraded without modifying the other modules.
\rephrase{For instance, supporting a new accelerator only requires updating the event handler, while users can add new tools without changing the handler or processor. This modular, extensible design makes \name{} suitable for diverse accelerators and targeted, low-overhead analysis.}

\begin{table}[t]
\caption{List of Supported Events in \name{}.} 
\label{tab:event_list}
\centering
\scriptsize
\begin{adjustbox}{width=0.4\textwidth}
\begin{tabular}{|c|c|l|}
\hline
\multirow{22}{*}{\rotatebox{90}{\textbf{\makecell{Low-Level\\Accelerator Events}}}} 
& \multirow{8}{*}{\textbf{\makecell{Coarse-Grained\\Host-Called\\API Events}}} & All Driver Functions \\
\cline{3-3}
& & All Runtime Functions \\
\cline{3-3}
& & Synchronization \\
\cline{3-3}
& & Kernel Launch \\
\cline{3-3}
& & Memory Copy \\
\cline{3-3}
& & Memory Set \\
\cline{3-3}
& & Resource Operations \\
\cline{3-3}
& & Batch Memory Operations \\
\cline{2-3}
& \multirow{15}{*}{\textbf{\makecell{Fine-Grained\\Device-Side\\Operations}}} & Thread Block Entry \\
\cline{3-3}
& & Thread Block Exit \\
\cline{3-3}
& & Global Memory Access \\
\cline{3-3}
& & Shared Memory Access \\
\cline{3-3}
& & Barrier Instruction \\
\cline{3-3}
& & Device Function Call \\
\cline{3-3}
& & Device Function Return \\
\cline{3-3}
& & Device-Side Malloc \\
\cline{3-3}
& & Device-Side Free \\
\cline{3-3}
& & Global-To-Shared Copy \\
\cline{3-3}
& & Pipeline Commit \\
\cline{3-3}
& & Pipeline Wait \\
\cline{3-3}
& & Remote Shared Memory Access \\
\cline{3-3}
& & Cluster Barrier \\
\cline{3-3}
& & Any Specific Instruction \\
\hline
\multicolumn{2}{|c|}{\multirow{7}{*}{\textbf{\makecell{High-Level\\DL Framework Events}}}}
& Operator Start \\
\cline{3-3}
\multicolumn{2}{|c|}{} & Operator End \\
\cline{3-3}
\multicolumn{2}{|c|}{} & Tensor Allocation \\
\cline{3-3}
\multicolumn{2}{|c|}{} & Tensor Reclamation \\
\cline{3-3}
\multicolumn{2}{|c|}{} & Layer Boundary* \\
\cline{3-3}
\multicolumn{2}{|c|}{} & Forward/Backward Boundary*\\
\cline{3-3}
\multicolumn{2}{|c|}{} & Customized Code Region*\\
\hline
\multicolumn{3}{l}{\tiny {* Requires manual insertion of \name{} annotations, as discussed in Section~\ref{range-analysis}.}}
\end{tabular}
\end{adjustbox}\vspace{-12pt}
\end{table}

\subsection{\name{} Modules}
\noindent\textbf{Event handler:}
In \name{}, the event handler module is responsible for initializing and setting up the profiling utilities. It abstracts the complexities of vendor-specific profiling APIs and DL framework callbacks, and provides a comprehensive set of handler functions for both coarse-grained and fine-grained runtime events on accelerators.
Coarse-grained events include kernel launches, memory copy operations, and synchronization calls, while fine-grained events capture 
individual thread-level activities, 
such as memory accesses by 
each thread.
In addition to low-level, “bare-metal” vendor-specific events, \name{} also monitors high-level DL framework-specific events, such as tensor allocation and operator execution.
Table~\ref{tab:event_list} summarizes the complete set of events currently supported by \name{}. 
\rephrase{\name{}’s extensible design allows new events to be supported by adding handler functions in the event handler module.}

\noindent\textbf{Event processor:} 
The event processor module pre-processes raw data captured by the event handler and dispatches it to the corresponding \name{} tool for customized analysis. It also normalizes event metadata across frameworks and profiling utilities, handling inconsistencies (e.g., negative vs. positive size for memory release) and extracting relevant details, such as grid configurations for kernel launch events or copy directions for memory copy operations, based on event type.

\begin{figure}[t]
\centering
\subfloat[Conventional GPU-based trace collection with CPU-side analysis. The GPU stalls when the trace buffer is full, waiting for the CPU to fetch and flush the data.\label{fig:cpu_analysis}]{
  \includegraphics[width=0.47\linewidth]{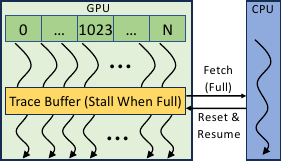}
}
\hfill
\subfloat[\name{}’s GPU-resident collect-and-analyze model. GPU threads perform in-situ analysis, avoiding stalls and reducing CPU-GPU overhead.\label{fig:gpu_analysis}]{
  \includegraphics[width=0.45\linewidth]{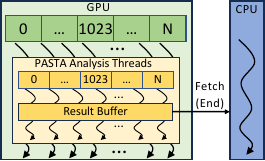}
}
\caption{Comparison of CPU- and GPU-based analysis models.}
\Description{}
\label{fig:cpu_gpu_analysis}\vspace{-10pt}
\end{figure}

\rephrase{To enable low-overhead analysis, the event processor adopts GPU-accelerated analysis by launching helper device functions that employ groups of GPU 
threads (e.g., warps in NVIDIA GPUs) to concurrently process collected data. When an event is triggered on the GPU (e.g., a memory access), the profiling library records the instruction into a device buffer. A helper device function then processes many of these events concurrently, significantly accelerating performance analysis.}
Figure~\ref{fig:cpu_gpu_analysis} compares CPU- and GPU-based analysis models. Specifically, Figure~\ref{fig:cpu_analysis} shows the conventional GPU-based trace collection with CPU-side analysis, which is used in vendor-provided tools such as the NVBit MemTrace tool~\cite{nvbit_samples} and the Compute Sanitizer MemoryTracker tool~\cite{compute_sanitizer_samples}. Figure~\ref{fig:gpu_analysis} demonstrates how \name{} adopts a GPU-resident collect-and-analyze model, effectively avoiding stalls and reducing CPU-GPU communication overhead.

\noindent\textbf{Tool collection:}
The collection module provides templates for customized analyses, and the user can retrieve all or a subset of events from the event processor module to conduct customized analyses for their own program analysis needs.
For example, they can extract tensor allocation and operator execution events to analyze memory usage or execution behavior in DL workloads, or retrieve kernel launch information to identify the most frequently invoked kernels.
We show several use cases in Section~\ref{sec:case-studies}.

\subsection{Workflow}
\begin{figure}[t]
    \centering
    \includegraphics[width=0.9\linewidth]{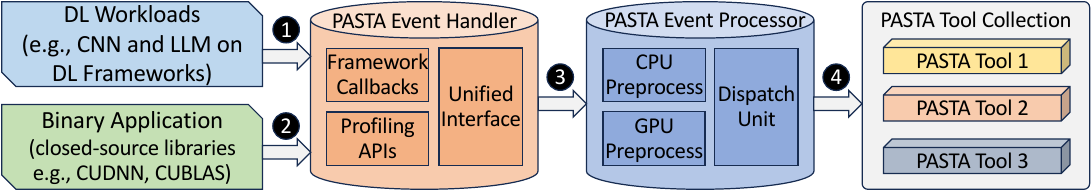}
    \caption{Workflow of \name{}.}
    \Description{}
    \label{fig:workflow}\vspace{-10pt}
\end{figure}

Figure~\ref{fig:workflow} shows the workflow of \name{}.
\name{} takes binary executable files of GPU-accelerated applications as input, without requiring access to the source code.
This makes \name{} particularly suitable for analyzing closed-source libraries such as CUDNN~\cite{nvidia-cudnn} and CUBLAS~\cite{nvidia-cublas}. During execution, when an event listed in Table~\ref{tab:event_list} occurs, the corresponding callback function in the \name{} event handler module is invoked (\ding{182} and \ding{183}), which collects meta and runtime information related to these events for subsequent processing. Once data collection is complete, the relevant function in the event processor pre-processes the gathered raw data (\ding{184}).
CPU preprocess functions handle coarse-grained events (e.g., memory allocations and kernel launches), while GPU preprocess functions manage fine-grained events (e.g., memory accesses).
Next, the dispatch unit routes the pre-processed data to a specific \name{} tool defined within the \name{} collection (\ding{185}). The selected tool then analyzes the data and generates performance reports about program behavior. Users can specify the desired \name{} tool via a command-line option or an environment variable, allowing flexible tool selection based on specific analysis.

\rev{\subsection{Support for Diverse GPU Platforms}}
\rev{To support GPUs from different vendors, \name{} provides a set of uniform, decoupled interfaces within its event handler. These interfaces simplify integration and ensure consistent event collection across platforms.}
For NVIDIA GPUs, \name{} takes advantage of callbacks from both the NVIDIA Compute Sanitizer APIs~\cite{compute-santizer-api} and NVIDIA NVBit~\cite{villa2019nvbit}.
The NVIDIA Compute Sanitizer APIs offer lightweight and 
intuitive 
callbacks that reduce 
development effort. 
\rephrase{However, they can only inspect a subset of instructions, such as memory and barrier operations.}
In contrast, NVIDIA NVBit offers more comprehensive coverage by covering all SASS instructions. 
This increased flexibility, however, requires 
substantial development effort and potentially incurs higher runtime overhead. 
\name{} allows users to have the flexibility to choose either of these libraries independently or use both in conjunction to gain insights into their code execution.
\rephrase{For AMD GPUs, \name{} integrates with the ROCprofiler-SDK tool library~\cite{rocprofiler-sdk}.
These APIs are analogous to NVIDIA’s Compute Sanitizer callbacks and enable \name{} to capture memory, kernel, and synchronization events on AMD platforms with the same interface. As a result, \name{} offers consistent cross-vendor support for profiling and analysis.}

\subsection{Support for Diverse DL Frameworks}
In DL frameworks, 
resources and GPU kernel executions are hierarchically managed, which often makes it challenging to adopt vendor-provided tools to gather insightful feedback
~\cite{min2021pytorch}. For example, in PyTorch and TensorFlow, GPU memory is managed via memory pools~\cite{cachingallocator}.
While pooled memories are first allocated via 
vendor-provided memory APIs (e.g., \texttt{cudaMalloc} or \texttt{HipMalloc}), subsequent allocations and releases 
of tensors are managed by memory pools that employ framework-specific memory management algorithms, which are often challenging to track synchronously with hardware events.
Furthermore, 
DL frameworks run 
one or multiple kernels within 
a single operator to 
complete a specific computation, where 
this operator-to-kernel mapping information is hidden 
from the users. 
To solve these challenges, \name{} 
leverages the callbacks~\cite{tensor-callback} provided by
DL frameworks to integrate high-level framework statistics
into the 
event handler module. 
Note that \name{} can collect low-level accelerator-related events and high-level DL framework-specific events concurrently, \rephrase{which fills the gap between} 
vendor-provided and DL-framework-provided profiling tools. 

\subsection{Advanced Features}

\begin{figure}[t]
\centering
\begin{adjustbox}{width=0.48\textwidth}
\begin{minipage}{\linewidth}
\begin{lstlisting}[caption={An example of layer-wise analysis support.
}, escapechar=|, label={lst:layerwise}]
|$\textcolor{ForestGreen}{\boldsymbol{+}}$|import pasta
# forward function of the model
def forward():
    ... # other layers
|$\textcolor{ForestGreen}{\boldsymbol{+}}$|    pasta.start() |\label{lst:start}|
    self.transformer_layer() # targeted region
|$\textcolor{ForestGreen}{\boldsymbol{+}}$|    pasta.stop() |\label{lst:end}|
\end{lstlisting}
\end{minipage}
\end{adjustbox}\vspace{-12pt}
\end{figure}

\subsubsection{Range-Specific Analysis}
\label{range-analysis}
It is common to analyze a specific sub-region of an application rather than the entire application. \name{} supports range-specific analysis to facilitate this need. For standard GPU applications, users can define the environment variables \texttt{START\_GRID\_ID} and \texttt{END\_GRID\_ID} to specify the subset of kernel launches to analyze. Additionally, \name{} provides support for Python annotations via the \texttt{pasta} package. 
Listing~\ref{lst:layerwise} shows an example usage of the \texttt{pasta} package. Users can annotate specific code regions they wish to analyze or profile using \texttt{pasta.start} and \texttt{pasta.end} (Lines~\ref{lst:start} and~\ref{lst:end}).

This feature is particularly useful in DL workloads, where 
individual layers typically have 
distinct behavioral characteristics. By leveraging this capability, users can conduct fine-grained analysis at the layer level, distinguish between forward and backward passes, or define any custom analysis range.

Although existing DL profiling tools offer similar annotation capabilities, \name{} distinguishes itself through its minimal and non-intrusive API design. Users can annotate regions of interest by simply inserting \texttt{pasta.start} and \texttt{pasta.end}, without needing to configure additional logging infrastructure or modify the execution context, 
enabling fine-grained performance analysis with minimal disruption to the original codebase.


\subsubsection{Inefficiency Location Utilities}

Identifying the source of performance inefficiencies is essential for effective optimization. \name{} provides cross-level location utilities that help developers pinpoint inefficient code at both high-level Python and low-level C/C++ levels, significantly simplifying the debugging and optimization process. In contrast, many existing analysis tools offer only partial visibility, such as low-level C/C++ backtraces (e.g., NVIDIA Nsight Systems~\cite{nvidia-systems}) or high-level Python call stacks (e.g., PyTorch Profiler~\cite{pytorch-profiler}), thus failing to deliver a comprehensive cross-level context for diagnosing inefficiencies.

\name{} enables selective control through a set of predefined knobs, such as \texttt{MAX\_MEM\_REFERENCED\_KERNEL} and \texttt{MAX\_CALLED\_KERNEL}, which identify the kernel with the most memory references and the most frequent invocations, respectively. Users can easily extend this mechanism with custom knobs to locate specific inefficiencies while avoiding the high overhead of capturing full context information for all runtime events.

\begin{figure}[t]
\centering
\includegraphics[width=0.95\linewidth]{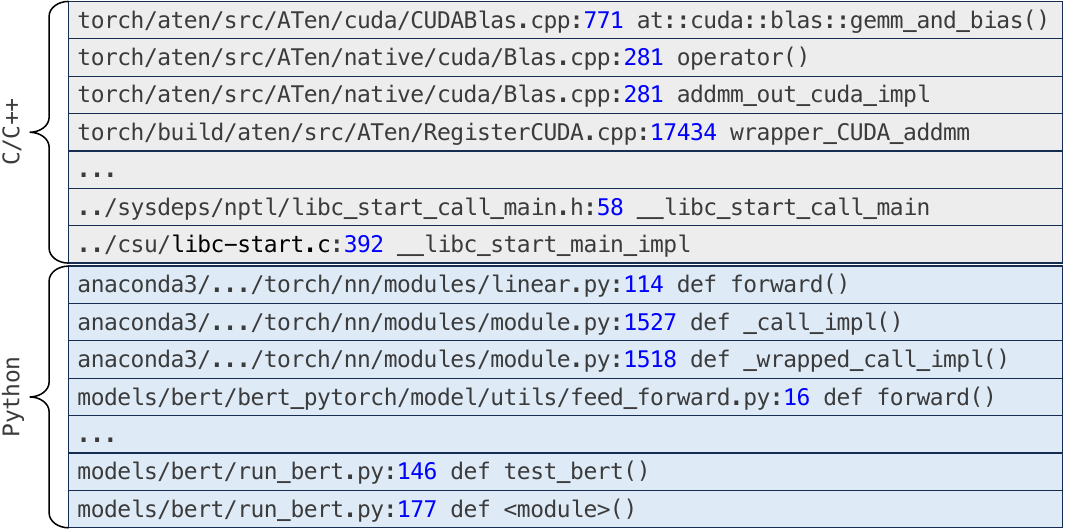}
\caption{Cross-layer call stack of the kernel with highest memory reference count during BERT inference. The trace spans Python-level code, PyTorch modules, and low-level C++/CUDA operations.}
\Description{}
\label{fig:callstacks}\vspace{-10pt}
\end{figure}

Figure~\ref{fig:callstacks} presents the call stack of the kernel with the highest memory reference count during BERT inference. This visualization enables users to easily identify the most memory-intensive kernel, \texttt{at::cuda::blas::gemm\_and\_bias}, facilitating targeted optimization for BERT execution on memory-bound systems.

\begin{figure*}[t]
    \centering
    \includegraphics[width=1\linewidth]{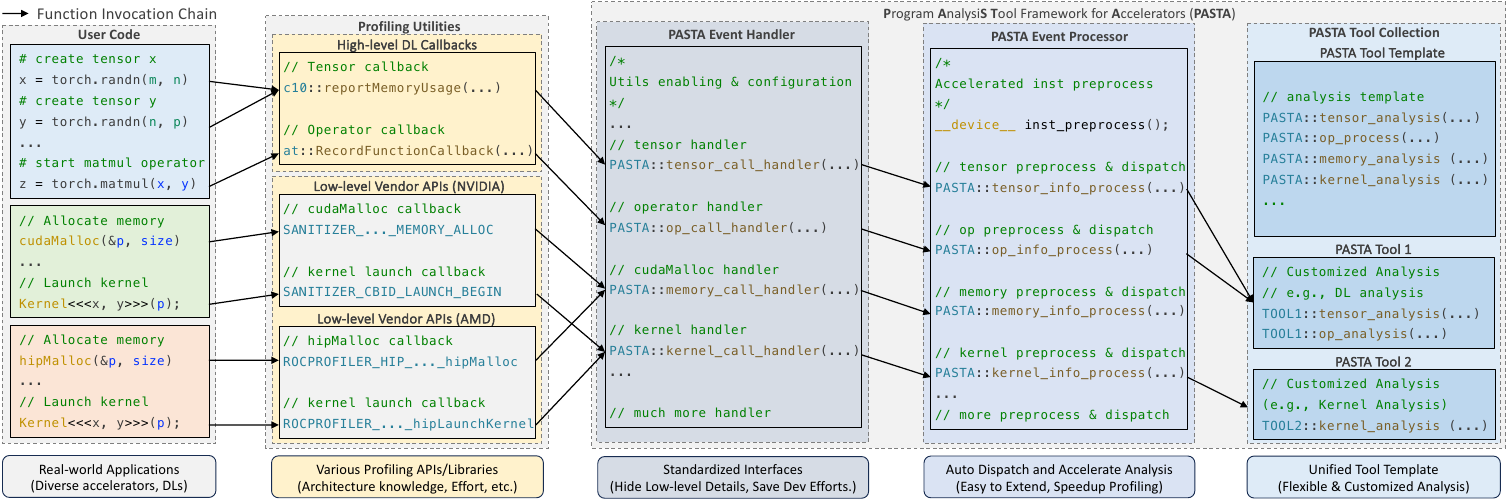}
        \caption{Codebase structure of \name{}.
        }
    \Description{}
    \label{fig:codebase}\vspace{-10pt}
\end{figure*}

\subsection{\rev{Generalization to Emerging Accelerators and Workloads}}
\rev{

\noindent\textbf{Support for Emerging Accelerators and Workloads.} 
\name{}'s architecture is designed to be adaptable beyond GPU-based accelerators and DL workloads.
\name{} can support accelerators if the accelerators provide runtime event instrumentation APIs, such as memory operations, kernel dispatch events, and synchronization points. Once the event APIs are provided, \name{} can be extended 
by implementing a backend handler that maps device-specific events (e.g., for Google TPUs, systolic-array operations or TPU counters) into \name{}’s unified event format.

Likewise, \name{} can also support workloads beyond DL because \name{} design is application agnostic. As far as the user specifies the region of interest based on his/her semantic knowledge of the target application, \name{} can be used for analyzing any applications, such as graph analytics or HPC applications. 

\noindent\textbf{Handling Differences in Low-Level Event Semantics.}
\name{} targets heterogeneous accelerators used as CPU co-processors.
Although terminologies may differ, many runtime events share common semantics: kernel launch events record grid size and kernel name, memory allocation events record address and size, and memory copy events specify size and transfer direction.
\name{}'s event handler normalizes such inconsistencies in event formats, naming conventions, and timing metadata. For instance, some runtimes report memory deallocation sizes with opposite signs or as deltas.
By abstracting such differences, \name{} unifies \textit{semantically equivalent} events and exposes a consistent interface to higher-level analyses.

Vendor-specific events, such as tensor memory operations in NVIDIA Blackwell GPUs or systolic-array operations in TPUs, are handled by specialized handler functions. These events are ignored on other accelerators, ensuring portability while exposing device-unique features.

}

\subsection{Extensibility \rephrase{for Diverse Analyses}}

\rephrase{The modular and unified architecture of \name{} makes it highly extensible for diverse analysis purposes. Developers can rapidly prototype instruction-level, memory-centric, or value-based tools with minimal changes, beyond the specific case studies in Section~\ref{sec:case-studies}.}

\noindent\textbf{Instruction-level analysis tools.}
These tools focus on fine-grained behaviors at the instruction granularity, leveraging \name{}’s support for instruction-level instrumentation via vendor APIs (as shown in Table~\ref{tab:event_list}).
\emph{Branch divergence analysis} can be implemented by intercepting device-side control flow instructions and correlating them with active thread masks, helping identify warp inefficiencies in SIMT architectures.
\emph{Instruction scheduling overhead analysis} targets pipeline stalls and issue port contention by analyzing throughput counters and stall reason metrics.
By integrating these with operator-level boundaries, developers can pinpoint inefficient scheduling regions.

\noindent\textbf{Memory-centric analysis tools.}
These tools examine how memory is used and accessed during execution, which is critical for understanding performance bottlenecks in memory-bound workloads.
\emph{Memory barrier stall analysis} quantifies synchronization delays that occur at device- or cluster-level barriers. With \name{}’s support for capturing barrier and synchronization events (as listed in Table~\ref{tab:event_list}), users can directly measure stall durations and frequencies. By recording timestamps at barrier entry and exit points, developers can compute precise stall intervals and identify kernels or layers that suffer from excessive synchronization overhead.
Additional analyses such as \emph{shared memory bank conflicts}, \emph{register pressure}, and \emph{underutilized memory regions} can be developed by leveraging \name{}’s memory event handler.

\noindent\textbf{Value-based analysis tools.}
These tools inspect runtime data values or semantics for correctness or anomaly detection.
For instance, a \emph{numeric overflow sanitizer} could instrument arithmetic instructions and track operand ranges to detect overflow or underflow events.
Similarly, tools such as \emph{redundant value load/store detection} and \emph{data taint tracking} can be implemented on top of \name{} by associating value semantics with traced instruction-level events. These analyses leverage \name{}'s fine-grained operation monitoring capabilities, such as operand values and memory accesses, to detect inefficiencies or security vulnerabilities during execution. 

%% file: implementation.tex
\section{Implementation
}\label{sec:impl}

As shown in Figure 5, the system is organized with five primary components: user code, profiling utilities, the event handler, the event processor, and custom analysis tools. 

\subsection{DL Supports}
\name{} integrates with real-world DL applications through both high-level and low-level interfaces. On the high-level side, it supports mainstream DL frameworks such as PyTorch via function hooks and callbacks (e.g., \texttt{c10::reportMemoryUsage} and \texttt{at::RecordFunction}).
At the low level, \name{} instruments accelerator-specific APIs.
For example, \name{} intercepts calls to \texttt{cudaMalloc} and \texttt{cuLaunchKernel} on NVIDIA platforms or \texttt{hipMalloc} and \texttt{hipLaunchKernel} on AMD platforms, providing fine-grained visibility into memory allocation and kernel launch events on the target hardware.

\subsection{\name{} Modules}
Figure~\ref{fig:codebase} presents the codebase structure of \name{}.
The event handler module receives diverse event information from both DL framework-level callbacks (e.g., \texttt{c10::reportMemoryUsage}) and low-level runtime instrumentation (e.g., \texttt{SANITIZER\_CBID\_LAUNCH\_BEGIN}), and translates them through a collection of modular handler functions (e.g., \texttt{PASTA::tensor\_call\_handler} for tensor allocations and \texttt{PASTA::kernel\_call\_handler} for kernel launches).
The event processor module then preprocesses the raw profiling data collected by the event handler using corresponding processor functions, such as \texttt{PASTA::tensor\_info\_process} and \texttt{PASTA::kernel\_info\_process}.
To support large volumes of fine-grained data—such as instruction-level access traces—\name{} employs GPU analysis threads via patched APIs (e.g., \texttt{sanitizerPatchModule}), accelerating preprocessing by offloading tasks to the device through \texttt{\_\_device\_\_}-annotated functions.
In the tool collection module, \name{} extracts relevant data for high-level analysis by overriding functions in customizable tool templates. 

\subsection{Interface to Target Application }
To enable seamless integration with target applications, \name{} is built as a shared library and injected at runtime via the \texttt{LD\_PRELOAD} mechanism. This allows it to intercept both framework and accelerator runtime calls without modifying application source code. Once loaded, \name{} enables the underlying event capture mechanisms using vendor-specific APIs. For instance, it utilizes \texttt{sanitizerEnableDomain} from Compute Sanitizer, \texttt{nvbit\_at\_cuda\_event} from NVBit, and \texttt{rocprofiler\_configure\_callback...} from the ROCProfiler SDK to enable and initialize the profiling utilities. Each captured event is handled via a corresponding callback implementation, which forwards the event to the event handler system. Finally, \name{} includes several advanced features to support rich developer introspection and cross-language analysis. The \texttt{pybind11} library is used to enable user annotations and tool customization via Python, while the \texttt{CPython PyFrame} API is leveraged to capture Python-level call stacks. For C/C++ sources, \name{} integrates with \texttt{libbacktrace} to extract symbolic stack traces.

\rev{
\subsection{Multi-GPU Support}

\name{} supports multi-GPU scenarios by associating events with the corresponding GPU using the device index exposed from vendor-provided profiling APIs~\cite{rocprofiler-sdk, compute-santizer-api}. 
Profiling multi-GPU computing has several challenges. One of them is the interference from auxiliary processes. To run an application on multiple GPUs, applications typically spawn one process to handle each GPU and use several helper processes~\cite{DistributedDataParallel, vllm-worker}. 
For instance, Megatron-LM~\cite{narayanan2021efficient} employs Just-In-Time (JIT) compilation that launches auxiliary processes during execution; when profiling with \texttt{LD\_PRELOAD}, these helpers—despite not creating a CUDA context—are still instrumented, leading to unnecessary initialization messages and potential runtime errors. To address this, \name{} uses \texttt{CUDA\_INJECTION64\_PATH} so that the profiler is injected only into processes that actually initialize a CUDA context. For multi-node GPU setups, \name{} runs independently on each node, generating profiles per rank or per node.
}

%% file: case-studies.tex
\section{Case Studies Using Tools Built with \name{}
}
\label{sec:case-studies}
In this section, we present several tools developed using \name{}, demonstrating how it aids developers in understanding performance issues and program behaviors, as well as guiding optimizations. \rev{Although we focus on DNN workloads in this paper, \name{} also supports other workloads, such as GPU-accelerated HPC applications.}

\input{evaluation}

\subsection{Common Application Behaviors Analysis}
In this subsection, we present two tools developed with \name{} that illustrate how users can extend \name{} 
\rephrase{for customized performance analysis.}
With only a few lines of code, they can analyze kernel invocation \rephrase{distributions} and identify optimization \rephrase{candidates}, showcasing the extensibility of \name{}. We also compare the profiling overhead of \name{} for different underlying profiling APIs and analysis mechanisms, highlighting \name{}’s flexible support for multiple profiling backends and its low-overhead design.

\begin{figure}[t]
\centering
\includegraphics[width=0.9\linewidth]{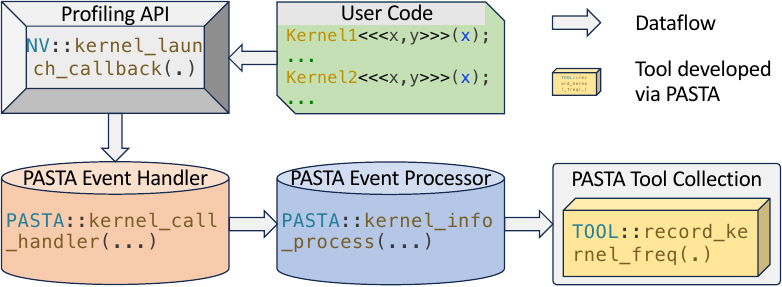}
\caption{
Kernel invocation 
analysis tool developed via \name{}.
}
\Description{}
\label{fig:kernel_analysis}\vspace{-10pt}
\end{figure}

\subsubsection{Kernel Invocation Frequency Analysis}
We first demonstrate a simple implementation of a kernel invocation frequency analysis tool to illustrate how \name{} can be extended for customized program analysis. We then present insights derived from the results of this analysis.

Figure~\ref{fig:kernel_analysis} shows the data flow of the kernel invocation frequency analysis tool. When a kernel launch event occurs, it triggers a kernel launch callback function provided by the profiling API (e.g., \texttt{NV::kernel\_launch\_callback}). This, in turn, invokes the \name{} event handler function \texttt{PASTA::kernel\_call\_handler}, which collects kernel-related information such as the kernel name and grid configuration. Subsequently, the \texttt{kernel\_info\_process} function in the event processor module preprocesses and organizes the data gathered by the event handler. These operations are handled entirely by the \name{} framework.

To develop a kernel invocation frequency analysis tool, users only need to retrieve this preprocessed data from the event processor and implement a customized analysis in the \texttt{TOOL::record\_kernel\_freq} function. In this case, users maintain a map to record the number of times each kernel is invoked—an intuitive yet insightful statistic.

\begin{figure}[t]
\centering
\includegraphics[width=0.95\linewidth]{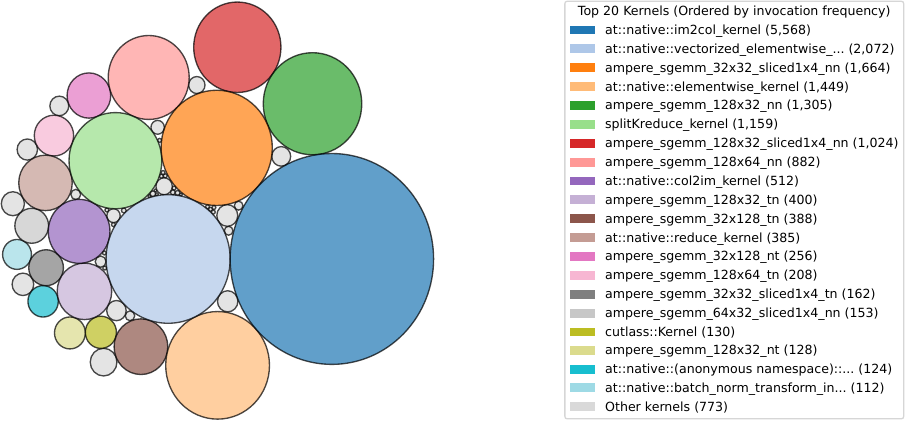}
\caption{Kernel invocation frequency distribution across all model inference and training runs: 
bubble size reflects invocation counts 
(actual numbers 
in the legend). 
}
\Description{}
\label{fig:all_models}\vspace{-10pt}
\end{figure}

Figure~\ref{fig:all_models} presents the kernel invocation frequencies observed during inference and training for the models listed in Table~\ref{tab:models}, as collected by the kernel frequency analysis tool. This analysis reveals several insights useful for optimization. Notably, although thousands of kernels are launched during model execution, only a small subset are invoked heavily—such as \texttt{at::native::im2col\_kernel} and \texttt{ampere\_sgemm*}. These results suggest that focusing optimization efforts on frequently invoked kernels can yield significant performance gains. 
Leveraging \name{}’s cross-layer call stack tracing feature, users can directly trace performance-critical kernels back to their source code, simplifying targeted optimizations, as shown in Figure~\ref{fig:callstacks}.
In comparison, existing tools that often require users to manually extract and correlate such patterns.

\begin{table}[t]
\caption{Memory characteristics of diverse DNN models (Sizes in MB unless otherwise noted).}
\label{tab:memory_result}
\centering
\scriptsize
\begin{adjustbox}{width=0.48\textwidth}
\begin{tabular}{|c|c|c|c|c|c|c|c|c|c|}
\hline
\multicolumn{2}{|c|}{Model} & \makecell{Kernel\\Count} & \makecell{Memory\\Footprint} & \makecell{Working\\Set (WS)} & \makecell{Minimum\\WS} & \makecell{Average\\WS} & \makecell{Median\\WS} & \makecell{90th\\percentile WS} \\
\hline
\hline
\multirow{6}{*}{\rotatebox{90}{Inference}} 
& AlexNet & 1428 & 1528.13 & 876.12 & 1.01 & 216.25 & 148.26 & 406.33 \\
\cline{2-9}
& RN-18 & 1497 & 1232.13 & 1024.0 & 1.00 KB & 86.07 & 64.00l & 172.27 \\
\cline{2-9}
& RN-34 & 2657 & 1261.59 & 1024.0 & 1.00 KB & 76.61 & 43.25 & 164.0 \\
\cline{2-9}
& BERT & 487 & 1179.64 & 212.62 & 47.50 KB & 75.23 & 37.69 & 141.75 \\
\cline{2-9}
& GPT-2 & 583 & 4148.10 & 1493.85 & 4.00 KB & 59.02 & 25.08 & 138.0 \\
\cline{2-9}
& Whisper & 663 & 2304.15 & 627.44 & 2.25 & 78.54 & 20.81 & 153.81 \\
\cline{2-9}
& Avg. & 1219 & 1942.29 & 876.34 & 0.55 & 98.62 & 56.52 & 196.03 \\
\hline
\hline
\multirow{6}{*}{\rotatebox{90}{Train}}
& AlexNet & 4040 & 3285.17 & 1512.09 & 512 B & 188.60 & 144.62 & 406.33 \\
\cline{2-9}
& RN-18 & 1542 & 3165.13 & 1024.00 & 512 B & 84.58 & 43.25 & 172.27 \\
\cline{2-9}
& RN-34 & 2734 & 4316.86 & 1024.00 & 512 B & 75.33 & 43.25 & 164.00 \\
\cline{2-9}
& BERT & 554 & 5679.03 & 235.47 & 1.00 KB & 77.71 & 37.97 & 209.30 \\
\cline{2-9}
& GPT-2 & 2004 & 7862.10 & 2240.77 & 512 B & 51.37 & 24.0 & 137.66 \\
\cline{2-9}
& Whisper & 665 & 2104.80 & 937.01 & 2.25 & 80.42 & 20.81 & 153.81 \\
\cline{2-9}
& Avg. & 2593 & 4402.02 & 1162.22 & 0.38 & 93.00 & 52.32 & 207.23 \\
\hline
\end{tabular}
\end{adjustbox}\vspace{-5pt}
\end{table}

\begin{figure}[t]
\centering
\subfloat[CPU-based analysis in conventional vendor-provided tools.\label{fig:cpu_memory_analysis}]{
\centering\includegraphics[width=0.9\linewidth]{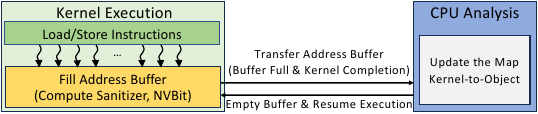}
}
\\
\subfloat[GPU-based analysis in \name{}.\label{fig:gpu_memory_analysis}]{
\centering\includegraphics[width=0.9\linewidth]{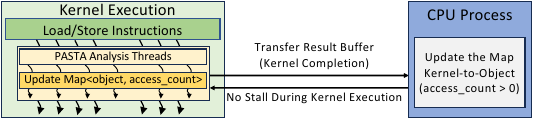}
}
\caption{
Memory characterization tool 
developed via \name{}.}
\Description{}
\label{fig:memory_analysis}\vspace{-10pt}
\end{figure} 

\subsubsection{Memory Characteristics Analysis}
\label{sec:memory_analysis_tool}
The memory characteristics analysis tool focuses on analyzing the \emph{working set size} of DL models. We define the working set size of a workload as the maximum memory footprint of any single kernel execution within that workload~\cite{working-set-size}. This metric is critical for evaluating whether the memory capacity of a system can accommodate a given workload.

However, analyzing the working set size of GPU-accelerated applications presents several challenges. First, existing profiling APIs, such as NVIDIA NVBit and AMD ROCProfiler SDK, only provide event-based metadata, such as kernel names and launch configurations, but not the argument lists or their values. This limitation makes it difficult to determine which memory objects are accessed by a given kernel. Second, even if the argument list is available, it is still possible that some objects passed into the kernel are never accessed, posing a challenge to accurately exclude them from the working set without tracking actual memory accesses.

To address these challenges, we developed a working set size analysis tool using \name{}. The core idea is to track which memory objects have been accessed during kernel execution. By associating memory access addresses with their corresponding objects, we can compute the memory footprint of each kernel. The maximum of these footprints across all kernels defines the working set size of the workload.

Table~\ref{tab:memory_result} summarizes the memory footprints and working set sizes for inference and training of the models listed in Table~\ref{tab:models}.
\rephrase{The results show: 1) working sets are often much smaller than overall footprints, with average footprints 2.22$\times$ and 3.79$\times$ larger than working sets in inference and training, respectively; 2) median and 90th percentile working sets are modest, indicating most kernels use limited memory. These findings suggest that a substantial fraction of memory is underutilized even for memory-intensive DL workloads. }
This insight provides theoretical support for memory optimization strategies such as swapping and data offloading~\cite{huang2020swapadvisor, ren2021zero, patil2022poet}.

\begin{figure}[t]
\centering
\includegraphics[width=0.9\linewidth]{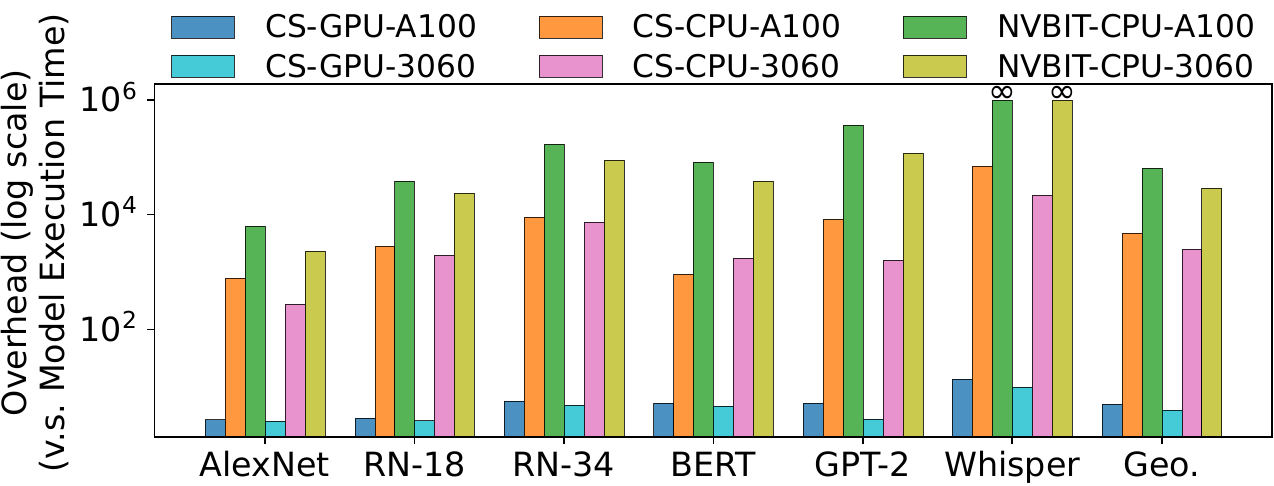}
\caption{Normalized overhead of diverse analysis models 
on A100 and RTX 3060. \mbox{CS-GPU}: 
GPU-side trace collection 
\& analysis using Compute Sanitizer. 
\mbox{CS-CPU}: 
trace collection on GPU \& 
analysis on CPU 
using Compute Sanitizer. 
\mbox{NVBIT-CPU}: 
trace collection on GPU \& 
analysis on CPU using NVBit. 
$\infty$ for those that did not finish within 7 days. 
}
\Description{}
\label{fig:overhead_comparison_all}\vspace{-8pt}
\end{figure}

\begin{figure}[t]
\centering
\includegraphics[width=0.9\linewidth]{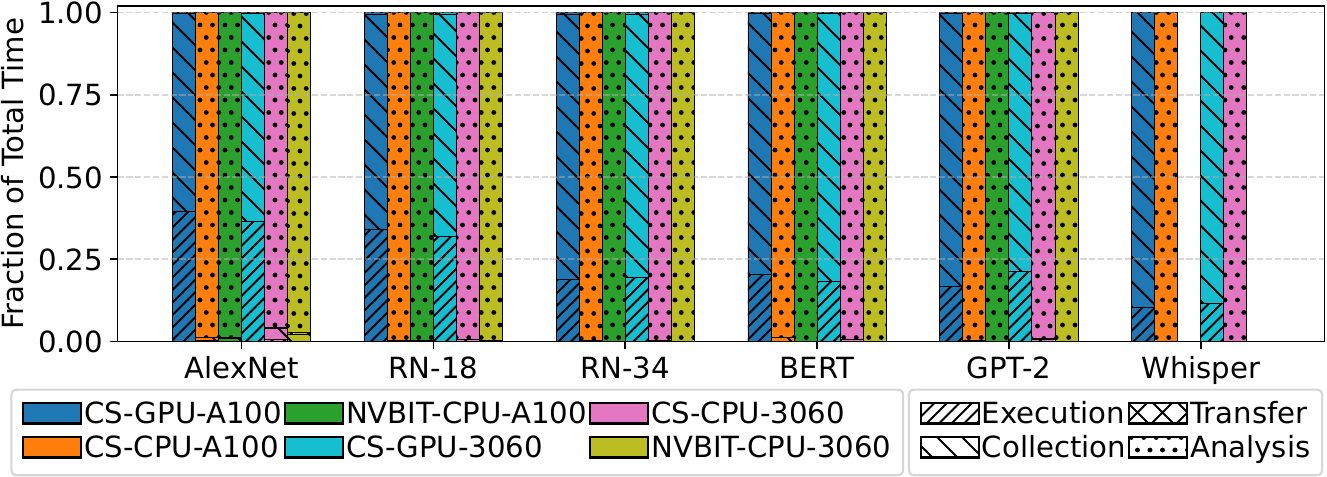}
\caption{\rev{Breakdown of \name{} profiling time on A100 and RTX 3060. (See Fig.~\ref{fig:overhead_comparison_all} for the definitions of CS-GPU, CS-CPU, NVBIT-CPU).}
}
\Description{}
\label{fig:overhead_breakdown}\vspace{-10pt}
\end{figure}

\subsubsection{Analysis Overhead of \name{}}
\label{sec:overhead}
As shown in Figure~\ref{fig:memory_analysis}, we implement the memory characteristics analysis tool described in Section~\ref{sec:memory_analysis_tool} in three variants: two conventional CPU-based approaches using Compute Sanitizer MemoryTracker tool~\cite{compute_sanitizer_samples} and NVBit MemTrace tool~\cite{nvbit_samples} correspondingly (Figure~\ref{fig:cpu_memory_analysis}), and another variant that uses GPU-accelerated analysis (Figure~\ref{fig:gpu_memory_analysis}).
In the CPU-based approaches, 
\rephrase{when memory instructions are instrumented, the log of accessed addresses is recorded into a buffer. The buffer is copied to the CPU when it becomes full or the kernel terminates, to be summarized for analysis. 
}
In contrast, the GPU-accelerated version performs this analysis directly on the device. When a kernel is launched, a map from memory object to access count is transferred to the GPU. During execution, a \hyeran{\rephrase{profiling device function increments access count for each associated memory object upon each access.}} 
When the kernel completes, the access count map is copied back to the CPU, where objects with non-zero access counts are identified as part of the kernel’s working set. 
\rephrase{By summarizing the profiling statistics on the device by exploiting GPU parallelism, this approach significantly accelerates analysis performance.} 

Figure~\ref{fig:overhead_comparison_all} compares the overhead of the GPU-accelerated analysis with the two CPU-based implementations on A100 GPUs and RTX 3060. On average, the results show that on A100, the GPU-accelerated tool in \name{} is 941$\times$ and 13006$\times$ faster than CPU-based tools using Compute Sanitizer and NVBit, respectively. On RTX 3060, it achieves average speedups of 627$\times$ and 7353$\times$. 
The CPU-based methods incur significant overhead as they rely on a single CPU thread and can introduce significant stalls. 
We also note that the Compute Sanitizer-based tool performs faster than the NVBit-based tool because it instruments only memory instructions, whereas NVBit must first dump and parse SASS code to identify memory instructions, which introduces additional overhead. 

\rev{We further break down the profiling overhead into four components: workload execution, trace collection, trace transfer, and trace analysis. Figure~\ref{fig:overhead_breakdown} shows the breakdown of \name{} profiling time on A100 and RTX 3060.
In the GPU-accelerated version, trace collection and analysis are fused into a single 
GPU 
function, so the reported ``collection time'' includes both collection and analysis.
Although collection time occupies a larger fraction in the GPU-accelerated version compared to CPU-based versions, its absolute time is much shorter, as shown in the overhead comparison in Figure~\ref{fig:overhead_comparison_all}. 
In contrast, CPU-based versions are dominated by trace analysis time, which could take 
hours to 
days since a limited number of (typically single) 
CPU threads process massive profiling data.}

\subsection{UVM Optimization for DL Workloads}
\label{sec:uvm-optimization}

\begin{figure}[t]
\centering
\includegraphics[width=0.9\linewidth]{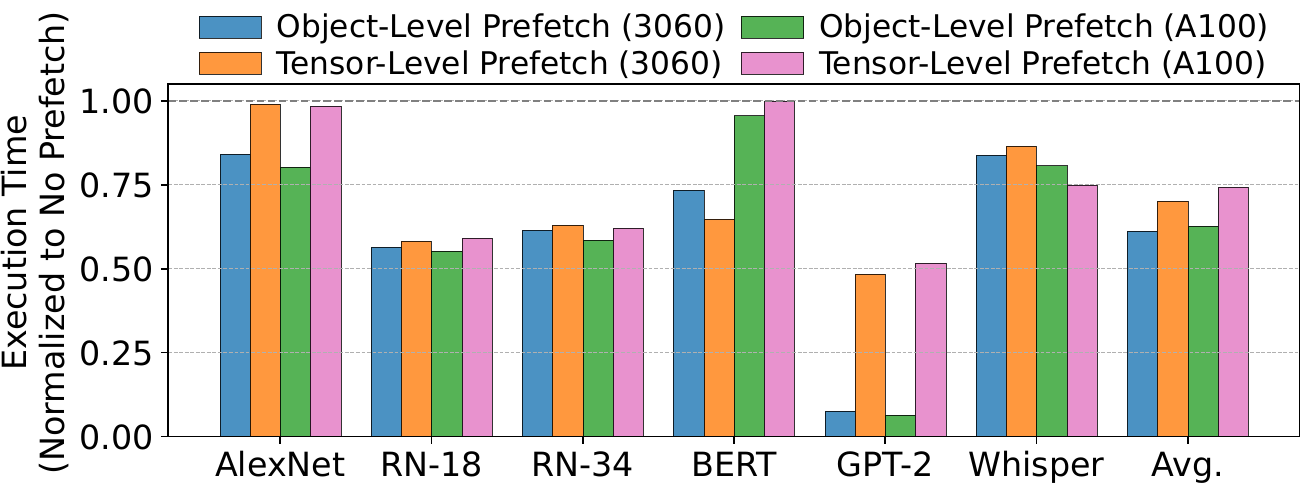}
\caption{Execution time of object-level and tensor-level prefetch on RTX 3060 and A100 under no memory oversubscription.}
\Description{}
\label{fig:uvm_speedup_all}\vspace{-10pt}
\end{figure}

\begin{figure}[t]
\centering
\includegraphics[width=0.9\linewidth]{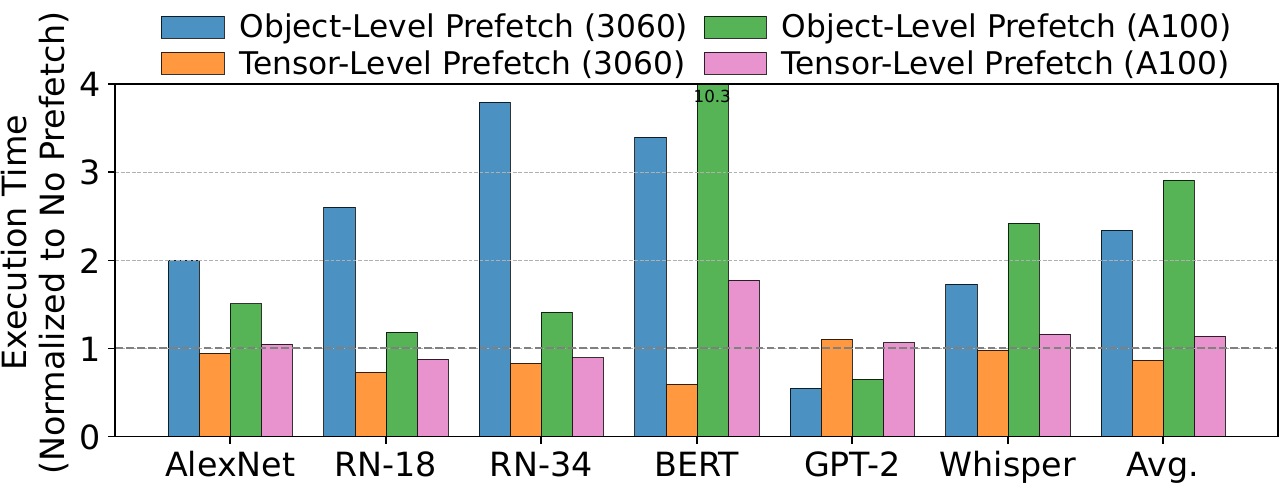}
\caption{Execution time of object-level and tensor-level prefetch on RTX 3060 and A100 under a memory oversubscription factor of 3.}
\Description{}
\label{fig:uvm_speedup_all_os}\vspace{-10pt}
\end{figure}

\begin{figure}[t]
\centering
\includegraphics[width=0.9\linewidth]{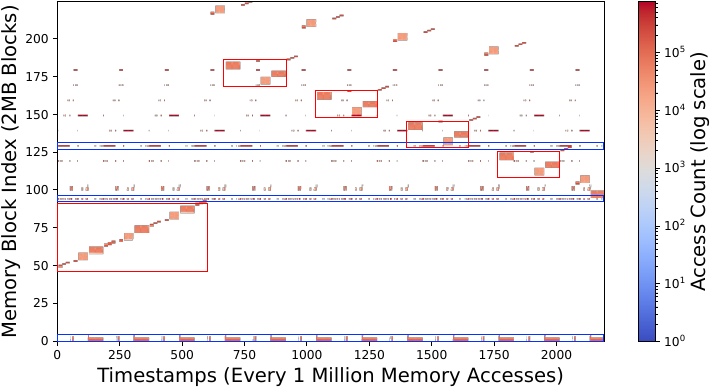}
\caption{Memory access hotness of BERT inference over time.}
\Description{}
\label{fig:bert_hotness}\vspace{-12pt}
\end{figure}

\subsubsection{Tensor-Aware UVM Prefetcher}
NVIDIA’s UVM provides a unified memory space shared between the GPU and CPU, simplifying GPU programming and enabling memory oversubscription to effectively expand the usable GPU memory.
Due to this advantage, UVM has been increasingly adopted for DL workloads~\cite{ramya2025vattention, lin2025understanding}, which have ever-growing memory demands~\cite{cMPI, CXL_performance, mLR, du2022fpga, ding2025vml}.
However, while UVM offers transparent memory expansion, its page-fault-driven, on-demand data migration mechanism can incur substantial overhead, especially when accessed data resides in CPU memory and must be migrated to the GPU at runtime~\cite{allen2021depth, lin2025understanding, go2023early}.

To mitigate these overheads, existing UVM optimization approaches aim to proactively prefetch or pre-evict data so that frequently accessed data resides in GPU memory, avoiding costly page fault handling~\cite{pratheek2024suv, lin2025forest}. These solutions typically operate at the granularity of memory objects (e.g., regions allocated via \texttt{cudaMallocManaged}), under the assumption that memory access patterns are consistent within each object. 
While this assumption holds for many conventional GPU applications, it \rephrase{does not apply to }
modern DL workloads. 
Contemporary DL frameworks such as PyTorch and TensorFlow adopt pool-based memory management. Instead of allocating memory per tensor, they request large chunks of memory from the system (using APIs like \texttt{cudaMalloc} or \texttt{cudaMallocManaged}) and then manage memory internally by subdividing these chunks into smaller regions to serve individual tensor allocations. As a result, a single memory object may contain multiple tensors, each with different lifetimes and access patterns.
This discrepancy renders existing object-level UVM prefetching strategies suboptimal for DL workloads~\cite{lin2025understanding}. Without awareness of tensor boundaries and usage patterns, object-level prefetching can result in unnecessary data migrations, memory bloat, and contention, thereby hurting performance.

To address this issue, we leverage \name{}’s cross-layer event capturing capability—capable of tracing both high-level framework-specific operations and low-level accelerator events—to develop a UVM prefetching analysis tool. This tool captures kernel execution events and correlates them with the accessed memory objects and tensors (as described in Section~\ref{sec:memory_analysis_tool}). Based on this analysis, we generate a multi-level prefetching scheme and build an automated UVM prefetcher that executes prefetching at either memory object or tensor granularity, and compares their performance.

Figure~\ref{fig:uvm_speedup_all} shows the normalized execution time of object-level and tensor-level prefetching on RTX 3060 and A100 GPUs under non-oversubscribed memory conditions. Both strategies yield improvements over the baseline (no prefetching), with average speedups of 39\% and 30\% on RTX 3060 and 37\% and 26\% on A100, respectively. Object-level prefetching achieves slightly higher speedups in this scenario, as it benefits from aggressive data migration when sufficient GPU memory is available.

However, under memory oversubscription, aggressive prefetching can be detrimental. Figure~\ref{fig:uvm_speedup_all_os} presents the normalized execution times under an oversubscription factor of 3 (i.e., the application’s memory footprint is 3× the GPU memory capacity).
In this case, object-level prefetching significantly degrades performance, with average slowdowns of 2.35$\times$ and 2.91$\times$ observed on RTX 3060 and A100, respectively. The root cause is that many tensors within a prefetched object may not actually be accessed during kernel execution, resulting in excessive and unnecessary data migration. This inefficient use of device memory leads to page thrashing and undermines performance. Notably, GPT-2 consistently benefits from object-level prefetching across both hardware platforms. This is attributed to its relatively small working set size compared to its overall memory footprint, as shown in Table~\ref{tab:memory_result}, which results in less memory pressure and minimal page thrashing, even under 3$\times$ oversubscription. While tensor-level prefetching outperforms the baseline on the RTX 3060, it performs slightly worse than the baseline on the A100. This highlights the need for more sophisticated prefetching strategies tailored to memory-intensive workloads, particularly when operating under constrained memory conditions.

\subsubsection{Time-Series Hotness Analysis}
The performance of UVM prefetching is determined by the timely delivery of ``hot'' data into the GPU. 
To research an efficient 
UVM prefetching algorithm, we develop a time-series hotness analysis tool using \name{}, which 
tracks access hotness over time in the unit of 2MB virtual memory blocks. Figure~\ref{fig:bert_hotness} shows the results 
of BERT inference without oversubscription. 
The results reveal significant divergence in access patterns across memory blocks. Memory blocks highlighted between each pair of the horizontal blue lines are frequently accessed throughout the entire execution, suggesting they likely store long-lived hot data (e.g., model parameters). 
These blocks are good candidates for prefetching and can be pinned in device memory using UVM APIs such as \texttt{cudaMemPrefetchAsync} and \texttt{cudaMemAdvise}.
In contrast, blocks highlighted with red boxes exhibit bursts of frequent accesses within narrow time windows and lack reusability, indicating they may contain short-lived, transient data (e.g., key-value caches). These blocks are suitable for proactive eviction to make room for other high-priority hot data.


\begin{figure}[t]
\centering
\includegraphics[width=0.9\linewidth]{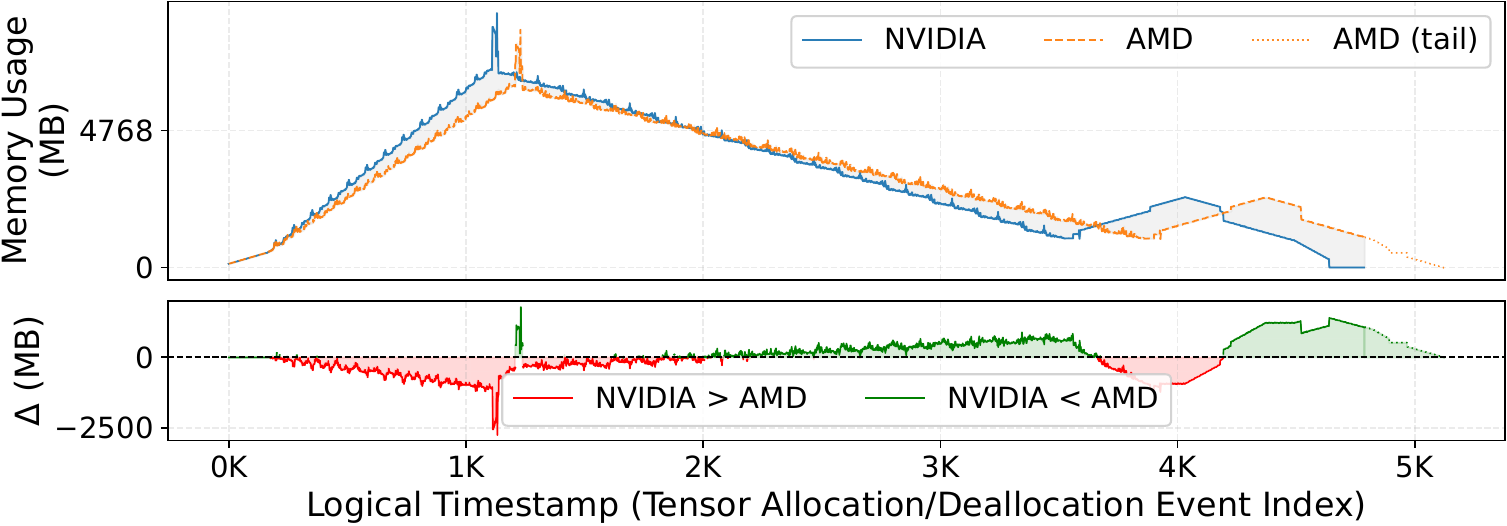}
\caption{\rev{Memory usage over time in one training iteration of GPT-2 under identical configurations on AMD and NVIDIA GPUs, with the bottom subfigure showing their difference.}}
\Description{}
\label{fig:amd_nvidia}\vspace{-12pt}
\end{figure}
\begin{figure}[t]
\centering
\subfloat[\rev{Data Parallelism}\label{fig:dp}]{
\centering\includegraphics[width=0.9\linewidth]{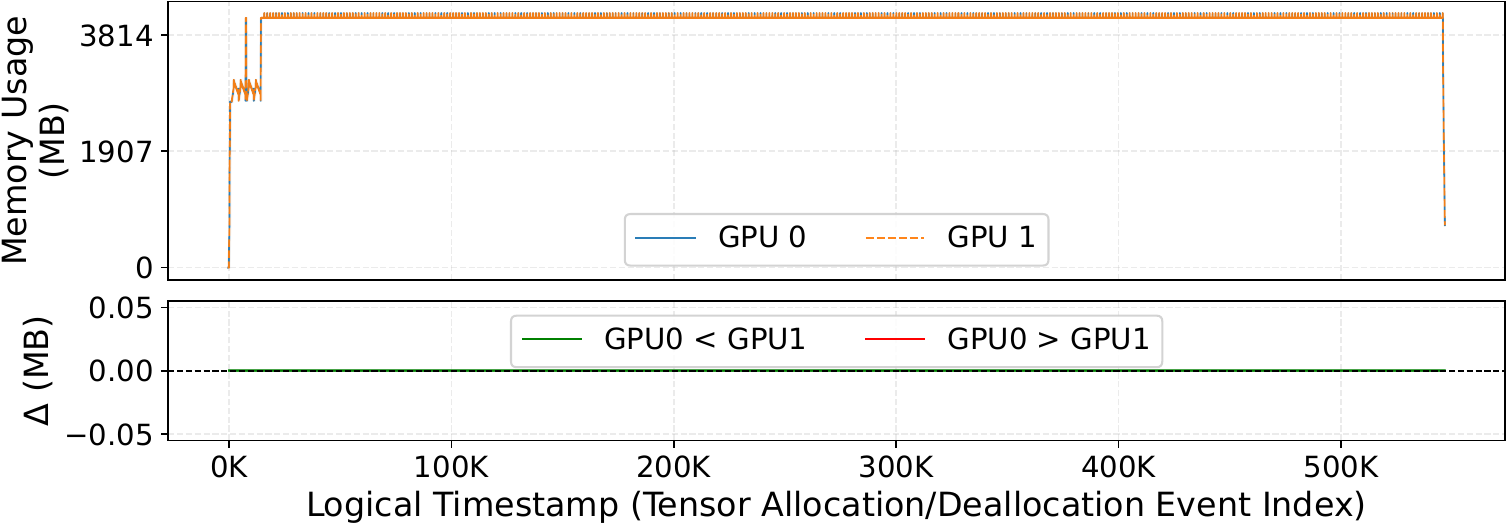}
}\vspace{-5pt}
\\
\subfloat[\rev{Tensor Parallelism}\label{fig:dp}]{
\centering\includegraphics[width=0.9\linewidth]{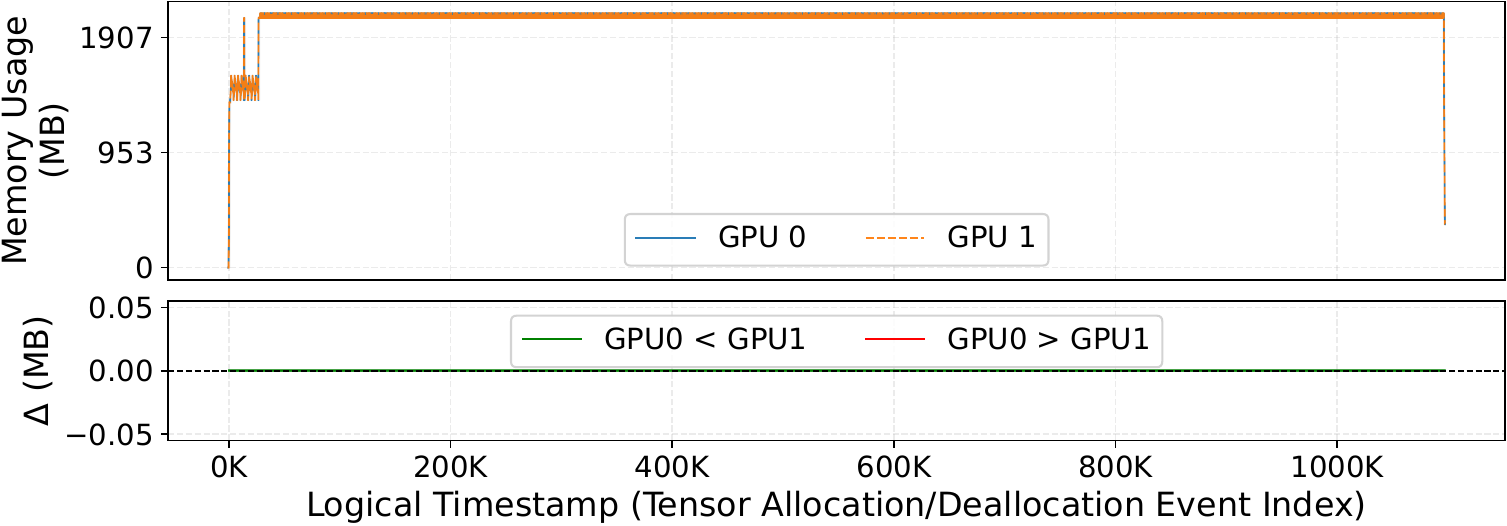}
}\vspace{-5pt}
\\
\subfloat[\rev{Pipeline Parallelism}\label{fig:dp}]{
\centering\includegraphics[width=0.9\linewidth]{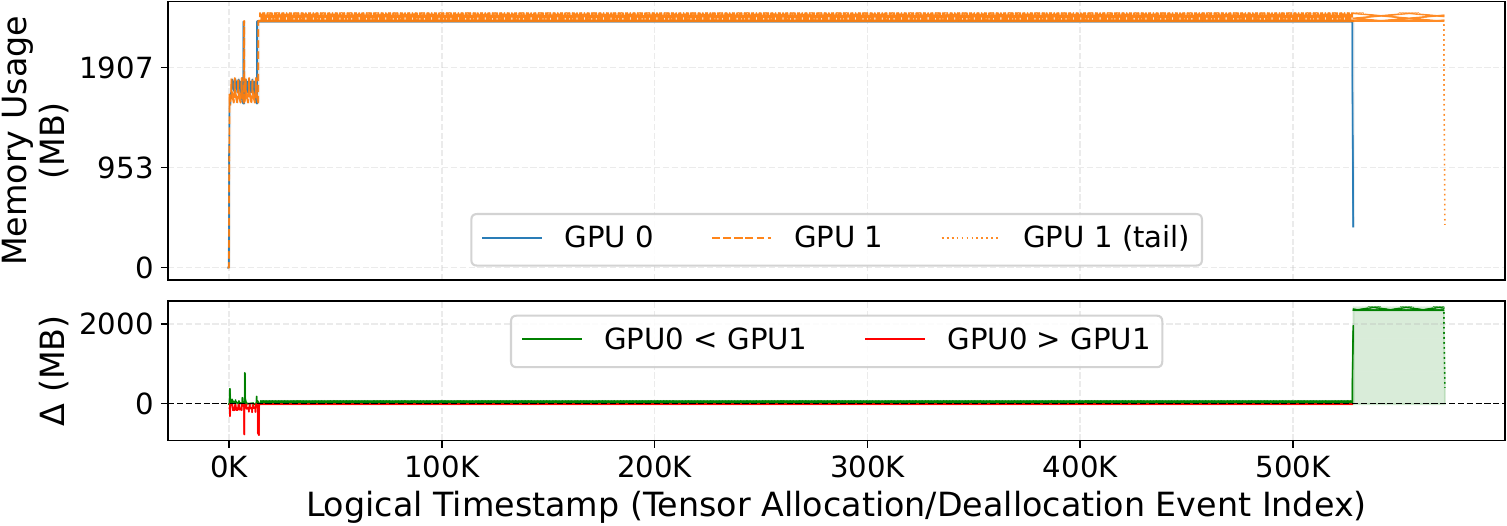}
}
\caption{\rev{Per-GPU memory usage over time in one training iteration of the Megatron GPT-2 345M model with different parallelism strategies. Bottom subfigures plot the memory usage difference between the two GPUs.}}
\Description{}
\label{fig:dist-training}\vspace{-12pt}
\end{figure} 

\rev{
\subsection{Support for Diverse GPU Vendors and Scenarios.}

\subsubsection{Comparison between AMD and NVIDIA GPUs}
\label{sec:amd-nvidia-gpu}
\name{} can support various GPU platforms. We compare the memory behaviors of NVIDIA and AMD GPUs (details in Table~\ref{tab:platform}) while running one training iteration of a GPT-2 model (Table~\ref{tab:models}). 
Figure~\ref{fig:amd_nvidia} shows the memory usage during the iteration. Both backends exhibit the same three-phase pattern—ramp-up, peak, ramp-down—as PyTorch’s caching allocator recycles tensors~\cite{pytorch-caching-allocator}. This similarity is expected since HIP memory management closely follows CUDA’s design~\cite{hip-torch-allocator}. We also observe backend-specific differences. On the NVIDIA GPU, fewer allocation/deallocation events are issued, but peak memory usage is slightly higher than on the AMD GPU. 
This discrepancy may be influenced by differences in operator decomposition and kernel fusion strategies across CUDA/cuDNN and HIP/MIOpen backends, as prior work has shown that fusion affects both the number of allocations and temporary memory requirements~\cite{zhang2024mcfuser, nie2021dnnfusion}.


\subsubsection{Multi-GPU Scenario}
We run Megatron GPT-2 345M~\cite{megatron-gpt2} on the Megatron-LM framework~\cite{narayanan2021efficient, megatron-lm} with two A100 GPUs (Table~\ref{tab:platform}). Figure~\ref{fig:dist-training} shows per-GPU memory usage over one training iteration under Data Parallelism (DP), Tensor Parallelism (TP), and Pipeline Parallelism (PP). Compared to the single-GPU case in Section~\ref{sec:amd-nvidia-gpu}, Megatron-LM’s memory behavior is different: tensors are more persistent with longer lifetimes (e.g., for communication). DP and TP exhibit identical memory usage across two GPUs, since DP runs two replicated models and TP evenly divides the model across devices. The peak memory of TP is about half of DP’s, consistent with model sharding. GPUs showed asymmetric statistics under PP because the model is split at the midpoint of the transformer block stack, thus final layers that produce logits run on GPU1, increasing GPU1’s tail execution.
These observations match the semantics of DP, TP, and PP, and demonstrate that \name{} can accurately reveal insights from complex workloads.
}

%% file: evaluation.tex
\subsection{Experimental Setup}
We evaluated the functionality and use cases of \name{} on \rev{three} CPU-GPU systems, \rev{each equipped with one or more discrete GPUs as accelerators}. Table~\ref{tab:platform} summarizes the hardware specifications and system software versions.

We studied six widely used DL models—AlexNet~\cite{krizhevsky2012imagenet}, ResNet18~\cite{he2016deep}, ResNet34~\cite{he2016deep}, GPT-2~\cite{radford2019language}, BERT~\cite{devlin2018bert}, and Whisper~\cite{radford2023robust}—as detailed in Table~\ref{tab:models}.
To control the UVM oversubscription factor (as applied in Section~\ref{sec:uvm-optimization}), we limit device memory capacity by allocating a specified amount in advance, following a common approach used in prior work~\cite{ganguly2020adaptive, ganguly2019interplay}.

\begin{table}[t]
\caption{Hardware and Software Environment.}
\label{tab:platform}
\centering
\scriptsize
\begin{adjustbox}{width=0.48\textwidth}
\begin{tabular}{|c|c|c|c|c|c|c|c|}
\hline
Machine & CPU & GPU & System & \makecell{System\\Memory} & \makecell{GPU\\Driver} & \makecell{\rev{GPU} \\Toolkit} \\
\hline
A & \makecell{Intel(R) Xeon(R)\\Gold 5320} & \makecell{NVIDIA\\A100 (80GB)} \rev{$\times 2$} & Linux 5.14 & 128 GB & 570.86.10 & \rev{CUDA} 12.1 \\
\hline
B & \makecell{AMD Ryzen 7\\5800X} & \makecell{NVIDIA GeForce\\RTX 3060} & Linux 6.11 & 32 GB & 560.28.03 & \rev{CUDA} 12.1 \\
\hline
\rev{C} & \rev{\makecell{Intel(R) Xeon(R)\\Platinum 8568Y}} & \rev{\makecell{AMD MI300X}} & \rev{Linux 6.8} & \rev{240 GB} & \rev{6.12.12} & \rev{ROCm 6.4} \\
\hline
\end{tabular}
\end{adjustbox}
\end{table}

\begin{table}[t]
\caption{Evaluated DL models.}
\label{tab:models}
\centering
\scriptsize
\begin{adjustbox}{width=0.4\textwidth}
\begin{tabular}{|c|c|c|c|c|c|c|}
\hline
Model & Type & Layers & Architecture & \makecell{Batch\\Size} & Abbr. \\
\hline
AlexNet & CNN & 8 & \makecell{Convolutional\\Full Connected} & 128 & AN\\
\hline
ResNet18 & CNN & 18 & \makecell{Residual Block} & 32 & RN-18\\
\hline
ResNet34 & CNN & 34 & \makecell{Residual Block} & 32 & RN-34\\
\hline
GPT-2 & Transformer  & 12 & \makecell{Transformer\\(Decoder)} & 8 & GPT-2\\
\hline
BERT & Transformer  & 12 & \makecell{Transformer\\(Encoder)} & 16 & BERT\\
\hline
\makecell{Whisper\\(small)} & Transformer  & 12 & \makecell{Transformer\\(En/De-coder)} & 16 & Whisper\\
\hline
\end{tabular}
\end{adjustbox}\vspace{-10pt}
\end{table}

%% file: discussion.tex
\rev{
\section{Discussion}

\subsection{Impact on Workload Execution.}
\noindent\textbf{Correctness.} 
\name{} passively intercepts runtime events but does not modify program data or execution logic. Thus, the functional correctness of the workload is unaffected, and all program outputs remain identical to uninstrumented execution. \name{} requires a small fraction of GPU memory (e.g., 4MB) to store profiling data. Therefore, 
\name{} induces minimal to no interference in resource usage.

\noindent\textbf{Performance Overhead.} 
\name{}'s runtime profiling may introduce performance overhead. The magnitude of this overhead is not strictly predictable, since it depends on both the type and volume of events being captured. In general, the more events or instructions are traced, the higher the expected overhead. \name{}’s GPU-accelerated design significantly mitigates these costs, as described in Section~\ref{sec:overhead}.

\subsection{Relation to Existing Techniques.}
\noindent\textbf{Stream Runtime Verification (SRV).}
SRV is an online verification mechanism that monitors the streams of events while an application is running and checks if the program executes as specified by the user~\cite{s-a-srv, bozzelli2014foundations}. 
\name{} leverages the runtime monitoring, similar to SRV. However, the goal and the overall mechanism of SRV and \name{} are fundamentally different.
SRV 
focuses on the program execution verification by using formalized specifications 
and various monitoring algorithms, whereas \name{}'s ultimate goal is to optimize program execution by providing accelerator-aware profiling APIs, tool templates, and backends that abstract vendor APIs. 

\noindent\textbf{eBPF-based Tracing.}
eBPF is widely used in Linux for dynamic tracing at the kernel level~\cite{gbadamosi2024ebpf}. It provides a programmable interface for collecting events such as system calls and I/O operations, enabling custom analysis tools. \name{} plays a comparable role for accelerators: it captures and normalizes GPU runtime events, offering modular templates for higher-level analysis. While eBPF addresses general-purpose observability, \name{} complements it by focusing on accelerator-specific semantics and GPU workloads.

}

%% file: conclusion.tex
\section{Conclusion}
In this paper, we present \name{}, a low-overhead, modular program analysis framework for heterogeneous accelerators. By unifying low-level profiling APIs with high-level framework callbacks, \name{} enables rapid development of customized analysis tools. Case studies on kernel invocation tracking, memory working set analysis, and UVM prefetch optimization demonstrate its versatility. Evaluations show that \name{} delivers significantly lower overhead than existing profilers while supporting rich cross-layer analysis, establishing its potential as a foundational tool for accelerator-aware system optimization and performance research.

\section*{Acknowledgment}
This work was supported by NSF grants, CAREER-2341039, CCF-2452081 and NSF-2411134. We thank AMD Cloud for providing computing resources.

%% file: appendix.tex
\section*{Artifact Appendix}

\subsection{Abstract}
Our artifact provides \name{}, a modular program analysis framework for accelerators, along with its profiling client AccelProf.
The artifact includes source code, build scripts, and detailed instructions to reproduce the main results presented 
in Figure~\ref{fig:all_models}, Table~\ref{tab:memory_result}, Figure~\ref{fig:overhead_comparison_all}, ~\ref{fig:overhead_breakdown}, ~\ref{fig:uvm_speedup_all}, ~\ref{fig:uvm_speedup_all_os}, ~\ref{fig:bert_hotness}, ~\ref{fig:amd_nvidia}, and ~\ref{fig:dist-training}.
The artifact demonstrates the case studies developed with \name{}.

\subsection{Artifact check-list (meta-information)}
{\small
\begin{itemize}
  \item {\bf Program: AccelProf.}
  \item {\bf Compilation: Makefile.}
  \item {\bf Run-time environment: Linux x86-64 systems.}
  \item {\bf Hardware: NVIDIA GPUs and AMD GPUs.}
  \item {\bf Execution: \texttt{accelprof -v -t <tool> <executable> [args...]}}
  \item {\bf Metrics: GPU application metrics demonstrating the functionality of the \name{} framework.}
  \item {\bf Output: Figures presented in the paper.}
  \item {\bf How much disk space required (approximately)?: $\leq$ 100 GB.}
  \item {\bf How much time is needed to prepare workflow (approximately)?: $\leq$ 1 hour.}
  \item {\bf How much time is needed to complete experiments (approximately)?: Reproducing Figure~\ref{fig:overhead_comparison_all} and Figure~\ref{fig:overhead_breakdown} may take several days. Other figures can be reproduced within $\leq$ 2 hour.}
  \item {\bf Publicly available?: Yes.}
  \item {\bf Code licenses (if publicly available)?: MIT.}
  \item {\bf Archived (provide DOI)?: \href{https://doi.org/10.5281/zenodo.17547322}{doi.org/10.5281/zenodo.17547322}}.
\end{itemize}

\subsection{Description}

\subsubsection{How delivered}
\normalsize
The artifact associated with this paper is publicly available at Zenodo~\cite{pasta_artifact}.

The open-source GitHub repository is publicly available at: \href{https://github.com/AccelProf/AccelProf}{https://github.com/AccelProf/AccelProf}.

User and developer documentation is publicly available at:
\href{https://accelprofdocs.readthedocs.io}{https://accelprofdocs.readthedocs.io}.

\subsubsection{Hardware dependencies}
\name{} supports both NVIDIA and AMD GPUs with x86-64 CPUs.
We have tested it on NVIDIA A100, NVIDIA GeForce RTX 3060, and AMD MI300X GPUs.
For best reproducibility, we recommend using the same GPU models and a machine with at least 100 GB of available disk space.

\subsubsection{Software dependencies}
The artifact was tested on the following software versions (or newer). Older versions may also work but are unverified.
\begin{itemize}
    \item NVIDIA CUDA driver: $\geq$ 560.28.03
    \item AMD GPU Driver $\geq$ 6.12.12
    \item CUDA Toolkit: 12.1 and above
    \item ROCm 6.4 and above
    \item GCC: 9.4 and above
    \item Linux Kernel: 5.14 and above
    \item PyTorch 2.0 and above
    \item NVIDIA NVBIT 1.7.3 and above
\end{itemize}

\subsection{Installation}
\begin{itemize}
\item Download the codebase.
The \name{} codebase is organized into multiple submodules.
\begin{lstlisting}[language=bash]
git clone --recursive \
    https://github.com/AccelProf/AccelProf.git
cd AccelProf && git checkout cgo26
git submodule update --init --recursive
\end{lstlisting}

\item Check dependencies
\name{} requires PyTorch and necessary Python development library installed.
\begin{lstlisting}[language=bash]
bash ./bin/utils/check_build_env.sh
\end{lstlisting}

\item Build \name{}.
\begin{lstlisting}[language=bash]
# 15 minutes
make ENABLE_CS=1 ENABLE_NVBIT=1 ENABLE_TORCH=1
\end{lstlisting}

\item Set environment variables.
\begin{lstlisting}[language=bash]
export ACCEL_PROF_HOME=$(pwd)
export PATH=${ACCEL_PROF_HOME}/bin:${PATH}
\end{lstlisting}

\item Setup \name{} AE toolkit.
\begin{lstlisting}[language=bash]
bash ./bin/setup_ae
\end{lstlisting}

\end{itemize}
\subsection{Experiment workflow}
\begin{itemize}
\item Setup artifact.
\begin{lstlisting}[language=bash]
cd cgo26-ae
bash ./bin/setup_artifact.sh
\end{lstlisting}

\item Reproduce Figure~\ref{fig:all_models}. Figure~\ref{fig:all_models} shows kernel invocation frequency distribution.
\begin{lstlisting}[language=bash]
bash ./bin/run_figure_7.sh
\end{lstlisting}

\item Reproduce Table~\ref{tab:memory_result}. Table~\ref{tab:memory_result} shows memory characteristics of diverse DNN models.
\begin{lstlisting}[language=bash]
bash ./bin/run_table_v.sh
\end{lstlisting}

\item Reproduce Figure~\ref{fig:overhead_comparison_all}.
Figure~\ref{fig:overhead_comparison_all} shows normalized overhead of diverse analysis models on A100 and RTX 3060.
This experiment may take several days to complete.
Users can set the environment variable \texttt{ACCEL_PROF_ENV_SAMPLE_RATE} to speed up the process.
\begin{lstlisting}[language=bash]
bash ./bin/run_figure_9.sh
\end{lstlisting}

\item Reproduce Figure~\ref{fig:overhead_breakdown}.
Figure~\ref{fig:overhead_breakdown} shows the breakdown of \name{} profiling time on A100 and RTX 3060.
This experiment may take several days to complete.
Users can set the environment variable \texttt{ACCEL_PROF_ENV_SAMPLE_RATE} to speed up the process.
\begin{lstlisting}[language=bash]
# Checkout to specific branch
cd ${ACCEL_PROF_HOME}
cd nv-nvbit && git checkout oh-breakdown
cd ${ACCEL_PROF_HOME}
cd nv-compute && git checkout oh-breakdown
cd ${ACCEL_PROF_HOME}

# Re-build the codebase
make ENABLE_CS=1 ENABLE_NVBIT=1 ENABLE_TORCH=1

# Run the experiment
# This may take several days to complete
bash ./bin/run_figure_10.sh
\end{lstlisting}

\item Reproduce Figure~\ref{fig:uvm_speedup_all}.
Figure~\ref{fig:uvm_speedup_all} shows execution time of object-level and tensor-level prefetch on RTX 3060 and A100 under no memory oversubscription.
\begin{lstlisting}[language=bash]
bash ./bin/run_figure_11.sh
\end{lstlisting}

\item Reproduce Figure~\ref{fig:uvm_speedup_all_os}.
Figure~\ref{fig:uvm_speedup_all_os} shows execution time of object-level and tensor-level prefetch on RTX 3060 and A100 under a memory oversubscription.

\begin{lstlisting}[language=bash]
bash ./bin/run_figure_12.sh
\end{lstlisting}

\item Reproduce Figure~\ref{fig:bert_hotness}.
Figure~\ref{fig:bert_hotness} shows memory access hotness of BERT inference over time.
\begin{lstlisting}[language=bash]
bash ./bin/run_figure_13.sh
\end{lstlisting}

\item Reproduce Figure~\ref{fig:amd_nvidia}.
Figure~\ref{fig:amd_nvidia} shows memory usage over time of GPT-2 training under identical configurations on AMD and NVIDIA GPUs.

Reproducing Figure 14 requires collecting data from both AMD and NVIDIA GPU platforms. The generated profiling trace from the AMD server must then be transferred to the NVIDIA server to plot the memory usage comparison.

\textbf{On AMD GPU:}
A file named \texttt{out_amd.log} will be generated in the \texttt{results/figure\_14/} directory. Please move this file to the corresponding \texttt{results/figure\_14/} directory on the NVIDIA server.
\begin{lstlisting}[language=bash]
# Download codebase
git clone --recursive \
    https://github.com/AccelProf/AccelProf.git
cd AccelProf && git checkout cgo26
git submodule update --init --recursive

# Compile the codebase
make ENABLE_ROCM=1

# Set environment ariables
export ACCEL_PROF_HOME=$(pwd)
export PATH=${ACCEL_PROF_HOME}/bin:${PATH}

# Setup AE Toolkit
bash bin/setup_ae
cd cgo26-ae
bash ./bin/setup_artifact.sh

# Run the experiment
bash ./bin/run_figure_14_amd.sh
\end{lstlisting}

\textbf{On NVIDIA GPU:}
After run the experiment for Figure 14 on AMD GPU, please move the \texttt{out_amd.log} to NVIDIA server under \texttt{results/figure _ 14/}.

\begin{lstlisting}[language=bash]
# Run the experiment
bash ./bin/run_figure_14_nvidia.sh

# Plot Figure 14
# Ensure out_amd.log is moved.
bash ./bin/plot_figure_14.sh results/figure_14/
\end{lstlisting}

\item Reproduce Figure~\ref{fig:dist-training}.
Figure~\ref{fig:dist-training} shows per-GPU memory usage over time in GPT-2 345M model training with different parallelism strategies.
The reproduction of Figure 15 requires Megatron-LM~\cite{narayanan2021efficient} to be installed.
\begin{lstlisting}[language=bash]
bash ./bin/run_figure_15.sh path_to_megatron
\end{lstlisting}

\end{itemize}

\subsection{Evaluation and expected result}
The reproduced results are located in folder \texttt{./results}.
The outputs for Figure~\ref{fig:all_models}, Table~\ref{tab:memory_result}, Figure~\ref{fig:overhead_comparison_all}, ~\ref{fig:overhead_breakdown}, ~\ref{fig:uvm_speedup_all}, ~\ref{fig:uvm_speedup_all_os}, ~\ref{fig:bert_hotness}, ~\ref{fig:amd_nvidia}, and ~\ref{fig:dist-training} are expected to match the corresponding results in the paper.



\subsection{Methodology}

Submission, reviewing and badging methodology:

\begin{itemize}
  \item \url{http://cTuning.org/ae/submission-20190109.html}
  \item \url{http://cTuning.org/ae/reviewing-20190109.html}
  \item \href{https://www.acm.org/publications/policies/artifact-review-badging}{https://www.acm.org/publications/policies/artifact-review-badging}
\end{itemize}